\documentclass{m}
\usepackage{graphicx}
\usepackage{hyperref}
\usepackage{xcolor}
\usepackage{amsmath}

\begin{document}

\def\bel#1{\begin{equation}\label{#1}}
\def\ee{\end{equation}}
\parindent=0pt
\parskip=0.35\baselineskip

\begin{frontmatter}
\title{Adsorption-desorption noise can be used for improving
selectivity}

\author{Alexander K.Vidybida}

\address{Bogolyubov Institute for Theoretical Physics\\
      Metrologichna str., 14-B, Kyiv 03143, Ukraine}

 \ead{vidybida@bitp.kiev.ua}
 \ead[url]{http://vidybida.kiev.ua}

\begin{abstract}
Small chemical sensors are subjected to adsorption-desorption fluctuations
which usually considered as noise contaminating useful signal. Based on
temporal properties of this noise, it is shown that it can be made useful if
proper processed. Namely,
the signal, which
characterizes the total amount of adsorbed analyte, should be subjected
to a kind of amplitude discrimination (or level crossing discrimination)
with
certain threshold. When the amount is equal or above the threshold, the result
of discrimination is standard dc signal, otherwise it is zero.
Analytes are applied at low concentration:
the mean adsorbed amount is below the threshold. The threshold is
 achieved  from time to time thanking to the fluctuations. The
signal after discrimination
 is averaged over a time window
 and used as the output of the whole
device.  Selectivity of this device is compared with that of its primary
adsorbing sites, based on explicit description of the threshold-crossing
statistics. It is concluded that the whole sensor may have much better
selectivity than do its individual adsorbing sites.
\end{abstract}
\end{frontmatter}

\section{Introduction}

Detectors of chemical substances are usually based on selective
adsorption-desorption (binding-releasing) of analyzed chemicals by
specific adsorbing sites (receptor mole\-cules).
 The receptor molecules are
attached to an electronic device able to measure the amount of the analyte
adsorbed during the binding-releasing process. The device may be either a MEMS
device, such as quartz crystal microbalance \cite{Lucklum,Snopok}, or
vibrating/bending cantilever \cite{Battiston}, or field effect transistor
\cite{Bartic}, or other \cite{Rittersma}. The device with the receptor
molecules is called chemical sensor or detector. In order to be useful, the
detector must be able to discriminate between different chemicals, to be
selective.  Its selectivity is normally the same as that of its receptor
molecules (see Eqs.(\ref{linprop},\ref{condetsel})).

The size of industrial sensors has constant tendency to decrease
\cite{Battiston}. The power of useful signal produced by a small detector
becomes very small. As a result, noise of the detector itself constitutes a
substantial portion of its output signal.
Depending on its construction, there are
several reasons for a small detector to be noisy \cite{VigKim}. One type of
noise is due to the fact that the adsorption-desorption process is driven by
brownian motion, which is stochastic. As a result, the instantaneous total
amount of adsorbed analyte is subjected to irregular fluctuations
visible in the output signal. This noise is called the
adsorption-desorption noise \cite{Yong1}. It is present in any small
detector which is based on binding-releasing of analyte. The
adsorption-desorption noise can dominate over all other types of intrinsic
noise \cite{Djuric}.

\begin{figure}[h]
\includegraphics[width=\textwidth]{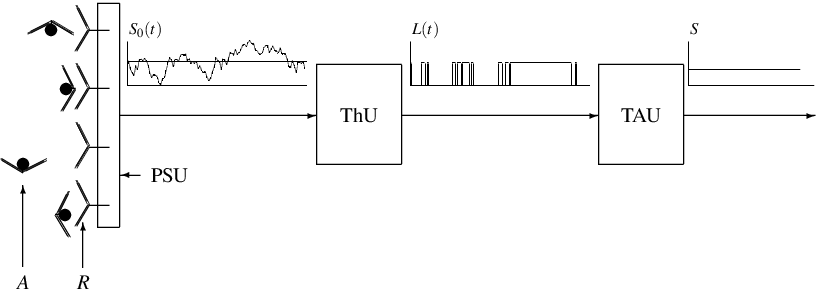}
\caption{\label{ThD}Schematic picture of threshold detector. A ---
analyte molecules; R --- adsorption sites; PSU --- primary
sensing unit; ThU --- threshold unit; TAU --- temporal averaging unit.}
\end{figure}

In this paper only the adsorption-desorption noise is taken into account. The
detector is expected to be a threshold detector (ThD), Fig.\ref{ThD}.

Namely, the fluctuating signal characterizing the amount of adsorbed analyte in
the primary sensing unit (PSU in Fig.\ref{ThD}) is fed into amplitude
discriminator unit (threshold unit,
ThU in Fig.\ref{ThD}).  The threshold unit is characterized by a certain
threshold. It has zero
as its output if the adsorbed amount is below the threshold, and it outputs
standard dc signal while the adsorbed amount is equal or above the
threshold.  The output of ThU is averaged over a sliding time-window to have
final output practically time-independent. This signal is considered
as the output of the ThD.

In this paper, the temporal properties of the binding-releasing stochastic
process are utilized to characterize the outputs of ThD if two analytes
are separately
presented at equal concentrations. This allows to compare selectivity of ThD
with that of its receptor molecules. The main conclusion is that the ThD may be
much more selective than do its adsorbing sites.

\section{Definitions and assumptions}

The adsorption-desorption process is described by the following
association-dissociation chemical reaction
\def\chh#1#2{\vcenter{\vbox{\hbox{ #1 }\hbox{ $\rightleftharpoons$
}\hbox{ #2 } } }}
\bel{chem1}
                 A+R \chh{$k_+$}{$k_-$} AR,
\ee
where $A$, $R$, $AR$ denote molecules of analyte, adsorption
site or receptor, and analyte-receptor
binary complex, respectively. At constant temperature, the rate constants,
$k_+$, $k_-$ are time-independent. They can be determined either from
experimental measurements, or estimated theoretically \cite{Djuric}. Let $N$
denotes the total number of receptor molecules per detector. The analyte is
presented at concentration $c$. The probability $p$ for any $R$ to be bound with
$A$ is\footnote{see \cite{VidBC}, where Eq.(\ref{proba1}) is justified.}
 \bel{proba1}
p={k_+c\over k_+c+k_-}.
\ee
The adsorption-desorption process is driven by brownian motion.
Therefore, instantaneous number of adsorbed molecules, $n(t)$,
changes in time randomly.
The mean number of adsorbed molecules, $\langle n\rangle$, can be calculated as
follows:
\bel{nmean}
\langle n\rangle=pN.
\ee
 If two different analytes $A_1$, $A_2$ are tested at the same concentration
 in two separate experiments,
either the Eq.(\ref{proba1}), or experimental measurements will give two
values, $p_1$, $p_2$. We say that the receptor molecule has selectivity with
respect to $A_1$, $A_2$, if $p_1\ne p_2$. We expect that
\bel{ineq}
p_1=p_2+\Delta p, \quad \Delta p >0.
\ee
The molecular
selectivity, $\mu$, is defined as
\footnote{\label{f1}This definition of selectivity differs
from one used in the previous version, namely, $\mu=\ln\frac{p_1}{p_2}$.
It can be easily checked that both definitions are equivalent for
small selectivities, when $\Delta p\to 0$.}
 \bel{mudef}
 \mu=\frac{\Delta p}{p_1}.
 \ee
 The primary signal, $S_0(t)$ in Fig.\ref{ThD}, usually increases
 if the number $n$ of adsorbed molecules increases:
 \bel{increasing}
 n>n'\Rightarrow S_0>S_0',
 \ee
 where the exact dependence of $S_0$ on $n$ is determined by the sensor
 construction and the transduction mechanism it employes.
 For simplicity, it is expected that in the case of gravimetric
 sensor, $A_1$ and $A_2$ have equal molecular masses.

 In the case of $A_1$ and $A_2$ used in the definition (\ref{mudef})
 one may expect that for the final output signal ($S$ in Fig.\ref{ThD})
 \bel{Sineq}
 S_1=S_2+\Delta S,\quad \Delta S>0,
 \ee
 where $S_1,S_2$ are the output levels if either $A_1$, or  $A_2$
 is applied.

 Define selectivity $\delta$ for a whole detector in terms of final output
 signal as follows:\footnote{See footnote \ref{f1}}
 \bel{defsel1}
 \delta=\frac{\Delta S}{S_1}.
 \ee

Both $S_0(t)$ and $n(t)$ are subjected to adsorption-desorption noise. In
 a detector without the threshold unit, the final output signal
 can be made linearly
 proportional to the mean number of adsorbed molecules:
 \bel{linprop}
 S_i\sim p_iN,\quad i=1,2.
 \ee
 This is achieved either by temporal averaging, or by
 choosing large detector with powerful primary signal in which contribution of
 adsorption-desorption fluctuations is not visible.  By using (\ref{mudef}),
 (\ref{linprop}) in (\ref{defsel1}) one obtains for selectivity of
 a conventional detector
\bel{condetsel}
 \delta=\frac{\Delta p N}{p_1 N}=\mu.
 \ee
 Thus, selectivity of detector in which the fluctuations are averaged out,
either
immediately after the primary sensing unit,
or inside it, equals to that of its individual
adsorbing sites.

 The threshold unit, ThU, rises a threshold which
 the $S_0$ must overcome
 in order to make possible further stages of processing. The crossing may
happen from time to time thanking to the adsorption-desorption fluctuations.
Due to
 (\ref{increasing}), the threshold can be characterized by the
 number $N_0$ of analyte molecules which must be adsorbed
 before the nonzero signal appears at the output end of the ThU.
 It is assumed that the ThU is ideal in a sense that the
 $N_0$ is the exact value which is not subjected to fluctuations.
 If $N_0$ is achieved, the ThU has standard constant signal as its output.
 The signal does not depend on the exact value of $n(t)$ provided it is above
 or equal to $N_0$.

 Denote by $T$ the temporal window over which the averaging is made in the TAU
 (Fig.\ref{ThD}), and by $T_b$, $T_a$ ($T_b+T_a=T$) the total amount of time
 the $n(t)$ spends below or above the threshold, respectively, when $0\le
 t\le T$. The final output, $S$ in Fig.\ref{ThD}, should be linearly
 proportional to $T_a/T$. From (\ref{ineq}), it is clear that
 $$
 T_{a1}=T_{a2}+\Delta T_a,
 $$
 where $T_{a1}$, $T_{a2}$ correspond to $A_1$, $A_2$, respectively.
 This gives for the selectivity of ThD:
 \bel{defsel2}
 \delta=\frac{\Delta T_a}{T_{a1}}.
 \ee

 \section{Estimation of selectivity gain}\label{estpar}

 Here we will introduce a selectivity gain $g$, which evaluates
 how much better selectivity of threshold detector can be if to compare
 with that of its individual receptor sites. The gain is defined
 as follows
 $$
 g = \frac{\delta}{\mu} = \frac{\Delta T_a\, p_1}{\Delta p\, T_{a1}}.
 $$
 For poor selectivities both $\Delta p$ and $\Delta T_a$ are small.
 Taking this into account the latter can be rewritten as a derivative:
 \bel{g2}
 g=\frac{p}{T_a}\frac{d\,T_a}{d\,p},
 \ee
 where $T_a$ is the amount of time spent above the threshold
 during period $T$ for a given  binding probability $p$ (we write here
 $p$ instead of $p_1$ to simplify expressions).

 It seems evident that
 \bel{TaProb}
 T_a=T\,\mathbf{Prob}\{n(t)\ge N_0\},
 \ee
 where
 \bel{Prob}
 \mathbf{Prob}\{n(t)\ge N_0\}=
 \sum\limits_{N_0\le k\le N}\binom{N}{k}p^k(1-p)^{N-k}.
 \ee
 Eq. (\ref{g2}), after substituting (\ref{TaProb}) and (\ref{Prob})
 turns ito the following:
$$
  g(p)=\frac{p
\sum\limits_{N_0\le k\le N} \frac{1}{k!(N-k)!}p^{k-1}(1-p)^{N-k-1}(k-Np)
  }{
  \sum\limits_{N_0\le k\le N} \frac{1}{k!(N-k)!}p^k(1-p)^{N-k}
  },
$$
which after simplification gives
\bel{gfin}
g(p)=\frac{
\sum\limits_{N_0\le k\le N} \frac{1}{k!(N-k)!}p^k(1-p)^{N-k}\frac{k-Np}{1-p}
  }{
  \sum\limits_{N_0\le k\le N} \frac{1}{k!(N-k)!}p^k(1-p)^{N-k}
}\,.
\ee
From (\ref{gfin}) it can be concluded that
\begin{equation}\label{ineq1}
    g(p)>\frac{N_0-Np}{1-p} = N\frac{p_0-p}{1-p},
\end{equation}
where $p_0=N_0/N$.
Taking into  account that the total number of adsorbing sites, $N$,
as well as $N_0$ can be
very large, it is clear from the estimate (\ref{ineq1}) that
selectivity gain $g$ can be
 much larger than $1$, provided the fraction $(p_0-p_1)/(1-p_1)$ is not very
 small. It must be at least positive, which requires
 \bel{ineq2}
 p_0>p_1,\quad\makebox{or }\quad p_1N<N_0.
 \ee
 Taking into account that $p$ increases with concentration (see
 Eq.(\ref{proba1})), inequality (\ref{ineq2}) can be considered as
 imposing an upper limit
 on concentration $c$ at which the effect of selectivity improvement might be
 expected based on the estimate (\ref{ineq1}). It is worth to notice that when
condition (\ref{ineq2}) holds,
the mean amount of adsorbed analyte is below the threshold
one, and threshold crossing may happen only due to fluctuations.

 \section{Numerical examples}

\begin{figure}
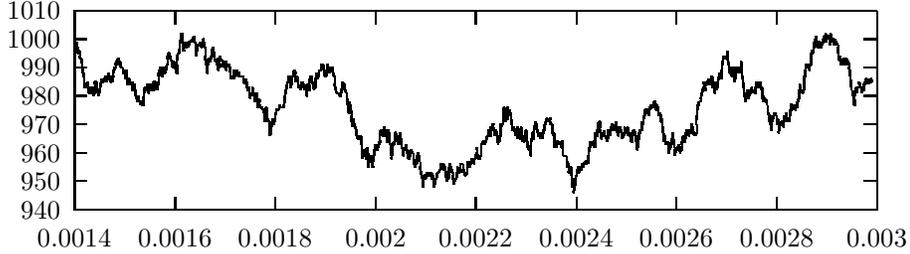

\setlength{\unitlength}{0.240900pt}
\ifx\plotpoint\undefined\newsavebox{\plotpoint}\fi
\sbox{\plotpoint}{\rule[-0.200pt]{0.400pt}{0.400pt}}%

\caption{\label{Trajec}Short segment of the trajectory $n(t)$ modelled
on PC for the Example 1 of Table \ref{Table}. Time (x-axis) is given in
seconds.  } \end{figure}

As one can conclude from the estimate (\ref{ineq1}), the selectivity
gain is higher for higher $N_0$. On the other hand, one cannot chose the
$N_0$ as high as desired because the ThU in Fig.\ref{ThD} is
expected to be ideal. If one
chose $N_0=100$ then the ideality means that the threshold level in the ThU is
allowed to have less then 1\%
jitter. Similarly, if one chose $N_0=10^4$ then the threshold level must be
kept with better than 0.01\%
precision. Otherwise, noise in the threshold level should be taken into account
in the reasoning of n.\ref{estpar}, and this will lead to a less promising
estimate.
\begin{table}[h]
\begin{center}
\begin{tabular}{ccc}
\hline
&$k_+$          &$k_-$\\
&(1/(s$\cdot$M))&(1/s)\\
\hline
$A_1$&1000&1000\\
$A_2$&1000&1050\\
\hline \end{tabular}
\end{center}
\caption{\label{Rates}
The rate constants used in the examples of Table \ref{Table} and in Fig.
\ref{Graphs}.
}
\end{table}
Another conclusion, based on the estimate (\ref{ineq1}), suggests that
the smaller is the
concentration (smaller $p$) of the analytes, the better is
discrimination between them. But in this case the threshold will be achieved
during small fraction of time spent for measuring. As a result, the
output signal will be very small and may be lost in the
TAU unit. It is natural to require that the output signal for more
affine analyte is higher than the 10\%
of the maximal output signal, which is produced if $n(t)\ge N_0$ all the time.
\begin{table}
\begin{center}
\begin{tabular}{ccccccc}
\hline
&$N$&$N_0$&$c$&$g$\\
&   &     &(M)&     \\
\hline
Example
1&10$^7$&10$^3$&9.6$\cdot10^{-5}$&72.6\\
Example 2&10$^8$&10$^4$&$9.9\cdot10^{-5}$
&360.0\\ \hline \end{tabular}\medskip
\end{center}
\caption{\label{Table}Numerical examples of improved selectivity.  The rate
constants for the analytes are shown in the Table \ref{Rates}.  $g$ is
calculated here by means of the exact expression (\ref{gfin}). }
\end{table}
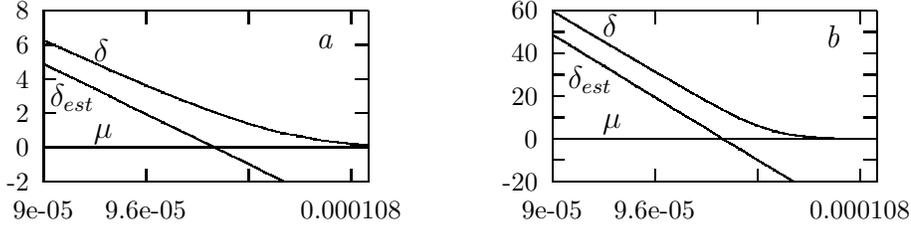
\begin{figure}
\setlength{\unitlength}{0.240900pt}
\ifx\plotpoint\undefined\newsavebox{\plotpoint}\fi
\sbox{\plotpoint}{\rule[-0.200pt]{0.400pt}{0.400pt}}%
\begin{picture}(750,360)(0,0)
\font\gnuplot=cmr10 at 10pt
\gnuplot
\sbox{\plotpoint}{\rule[-0.200pt]{0.400pt}{0.400pt}}%
\put(176.0,122.0){\rule[-0.200pt]{122.859pt}{0.400pt}}
\put(176.0,68.0){\rule[-0.200pt]{4.818pt}{0.400pt}}
\put(154,68){\makebox(0,0)[r]{-2}}
\put(666.0,68.0){\rule[-0.200pt]{4.818pt}{0.400pt}}
\put(176.0,122.0){\rule[-0.200pt]{4.818pt}{0.400pt}}
\put(154,122){\makebox(0,0)[r]{0}}
\put(666.0,122.0){\rule[-0.200pt]{4.818pt}{0.400pt}}
\put(176.0,176.0){\rule[-0.200pt]{4.818pt}{0.400pt}}
\put(154,176){\makebox(0,0)[r]{2}}
\put(666.0,176.0){\rule[-0.200pt]{4.818pt}{0.400pt}}
\put(176.0,229.0){\rule[-0.200pt]{4.818pt}{0.400pt}}
\put(154,229){\makebox(0,0)[r]{4}}
\put(666.0,229.0){\rule[-0.200pt]{4.818pt}{0.400pt}}
\put(176.0,283.0){\rule[-0.200pt]{4.818pt}{0.400pt}}
\put(154,283){\makebox(0,0)[r]{6}}
\put(666.0,283.0){\rule[-0.200pt]{4.818pt}{0.400pt}}
\put(176.0,337.0){\rule[-0.200pt]{4.818pt}{0.400pt}}
\put(154,337){\makebox(0,0)[r]{8}}
\put(666.0,337.0){\rule[-0.200pt]{4.818pt}{0.400pt}}
\put(176.0,68.0){\rule[-0.200pt]{0.400pt}{4.818pt}}
\put(176,23){\makebox(0,0){9e-05}}
\put(176.0,317.0){\rule[-0.200pt]{0.400pt}{4.818pt}}
\put(337.0,68.0){\rule[-0.200pt]{0.400pt}{4.818pt}}
\put(337,23){\makebox(0,0){9.6e-05}}
\put(337.0,317.0){\rule[-0.200pt]{0.400pt}{4.818pt}}
\put(498.0,68.0){\rule[-0.200pt]{0.400pt}{4.818pt}}
\put(498.0,317.0){\rule[-0.200pt]{0.400pt}{4.818pt}}
\put(659.0,68.0){\rule[-0.200pt]{0.400pt}{4.818pt}}
\put(659,23){\makebox(0,0){0.000108}}
\put(659.0,317.0){\rule[-0.200pt]{0.400pt}{4.818pt}}
\put(176.0,68.0){\rule[-0.200pt]{122.859pt}{0.400pt}}
\put(686.0,68.0){\rule[-0.200pt]{0.400pt}{64.802pt}}
\put(176.0,337.0){\rule[-0.200pt]{122.859pt}{0.400pt}}
\put(256,143){\makebox(0,0)[l]{$\mu$}}
\put(256,200){\makebox(0,0)[r]{$\delta_{est}$}}
\put(256,274){\makebox(0,0)[l]{$\delta$}}
\put(605,289){\makebox(0,0)[l]{\it a}}
\put(176.0,68.0){\rule[-0.200pt]{0.400pt}{64.802pt}}
\put(176,123){\usebox{\plotpoint}}
\put(176.0,123.0){\rule[-0.200pt]{122.859pt}{0.400pt}}
\put(176,290){\usebox{\plotpoint}}
\multiput(176.00,288.92)(1.142,-0.492){21}{\rule{1.000pt}{0.119pt}}
\multiput(176.00,289.17)(24.924,-12.000){2}{\rule{0.500pt}{0.400pt}}
\multiput(203.00,276.92)(1.142,-0.492){21}{\rule{1.000pt}{0.119pt}}
\multiput(203.00,277.17)(24.924,-12.000){2}{\rule{0.500pt}{0.400pt}}
\multiput(230.00,264.92)(1.142,-0.492){21}{\rule{1.000pt}{0.119pt}}
\multiput(230.00,265.17)(24.924,-12.000){2}{\rule{0.500pt}{0.400pt}}
\multiput(257.00,252.92)(1.099,-0.492){21}{\rule{0.967pt}{0.119pt}}
\multiput(257.00,253.17)(23.994,-12.000){2}{\rule{0.483pt}{0.400pt}}
\multiput(283.00,240.92)(1.251,-0.492){19}{\rule{1.082pt}{0.118pt}}
\multiput(283.00,241.17)(24.755,-11.000){2}{\rule{0.541pt}{0.400pt}}
\multiput(310.00,229.92)(1.142,-0.492){21}{\rule{1.000pt}{0.119pt}}
\multiput(310.00,230.17)(24.924,-12.000){2}{\rule{0.500pt}{0.400pt}}
\multiput(337.00,217.92)(1.251,-0.492){19}{\rule{1.082pt}{0.118pt}}
\multiput(337.00,218.17)(24.755,-11.000){2}{\rule{0.541pt}{0.400pt}}
\multiput(364.00,206.92)(1.251,-0.492){19}{\rule{1.082pt}{0.118pt}}
\multiput(364.00,207.17)(24.755,-11.000){2}{\rule{0.541pt}{0.400pt}}
\multiput(391.00,195.92)(1.381,-0.491){17}{\rule{1.180pt}{0.118pt}}
\multiput(391.00,196.17)(24.551,-10.000){2}{\rule{0.590pt}{0.400pt}}
\multiput(418.00,185.92)(1.329,-0.491){17}{\rule{1.140pt}{0.118pt}}
\multiput(418.00,186.17)(23.634,-10.000){2}{\rule{0.570pt}{0.400pt}}
\multiput(444.00,175.93)(1.543,-0.489){15}{\rule{1.300pt}{0.118pt}}
\multiput(444.00,176.17)(24.302,-9.000){2}{\rule{0.650pt}{0.400pt}}
\multiput(471.00,166.93)(1.543,-0.489){15}{\rule{1.300pt}{0.118pt}}
\multiput(471.00,167.17)(24.302,-9.000){2}{\rule{0.650pt}{0.400pt}}
\multiput(498.00,157.93)(1.748,-0.488){13}{\rule{1.450pt}{0.117pt}}
\multiput(498.00,158.17)(23.990,-8.000){2}{\rule{0.725pt}{0.400pt}}
\multiput(525.00,149.93)(2.018,-0.485){11}{\rule{1.643pt}{0.117pt}}
\multiput(525.00,150.17)(23.590,-7.000){2}{\rule{0.821pt}{0.400pt}}
\multiput(552.00,142.93)(2.936,-0.477){7}{\rule{2.260pt}{0.115pt}}
\multiput(552.00,143.17)(22.309,-5.000){2}{\rule{1.130pt}{0.400pt}}
\multiput(579.00,137.93)(2.825,-0.477){7}{\rule{2.180pt}{0.115pt}}
\multiput(579.00,138.17)(21.475,-5.000){2}{\rule{1.090pt}{0.400pt}}
\multiput(605.00,132.94)(3.844,-0.468){5}{\rule{2.800pt}{0.113pt}}
\multiput(605.00,133.17)(21.188,-4.000){2}{\rule{1.400pt}{0.400pt}}
\multiput(632.00,128.95)(5.820,-0.447){3}{\rule{3.700pt}{0.108pt}}
\multiput(632.00,129.17)(19.320,-3.000){2}{\rule{1.850pt}{0.400pt}}
\put(659,125.17){\rule{5.500pt}{0.400pt}}
\multiput(659.00,126.17)(15.584,-2.000){2}{\rule{2.750pt}{0.400pt}}
\put(176,253){\usebox{\plotpoint}}
\multiput(176.00,251.92)(1.052,-0.493){23}{\rule{0.931pt}{0.119pt}}
\multiput(176.00,252.17)(25.068,-13.000){2}{\rule{0.465pt}{0.400pt}}
\multiput(203.00,238.92)(1.052,-0.493){23}{\rule{0.931pt}{0.119pt}}
\multiput(203.00,239.17)(25.068,-13.000){2}{\rule{0.465pt}{0.400pt}}
\multiput(230.00,225.92)(1.052,-0.493){23}{\rule{0.931pt}{0.119pt}}
\multiput(230.00,226.17)(25.068,-13.000){2}{\rule{0.465pt}{0.400pt}}
\multiput(257.00,212.92)(1.012,-0.493){23}{\rule{0.900pt}{0.119pt}}
\multiput(257.00,213.17)(24.132,-13.000){2}{\rule{0.450pt}{0.400pt}}
\multiput(283.00,199.92)(0.974,-0.494){25}{\rule{0.871pt}{0.119pt}}
\multiput(283.00,200.17)(25.191,-14.000){2}{\rule{0.436pt}{0.400pt}}
\multiput(310.00,185.92)(1.052,-0.493){23}{\rule{0.931pt}{0.119pt}}
\multiput(310.00,186.17)(25.068,-13.000){2}{\rule{0.465pt}{0.400pt}}
\multiput(337.00,172.92)(1.052,-0.493){23}{\rule{0.931pt}{0.119pt}}
\multiput(337.00,173.17)(25.068,-13.000){2}{\rule{0.465pt}{0.400pt}}
\multiput(364.00,159.92)(1.052,-0.493){23}{\rule{0.931pt}{0.119pt}}
\multiput(364.00,160.17)(25.068,-13.000){2}{\rule{0.465pt}{0.400pt}}
\multiput(391.00,146.92)(1.052,-0.493){23}{\rule{0.931pt}{0.119pt}}
\multiput(391.00,147.17)(25.068,-13.000){2}{\rule{0.465pt}{0.400pt}}
\multiput(418.00,133.92)(1.012,-0.493){23}{\rule{0.900pt}{0.119pt}}
\multiput(418.00,134.17)(24.132,-13.000){2}{\rule{0.450pt}{0.400pt}}
\multiput(444.00,120.92)(1.052,-0.493){23}{\rule{0.931pt}{0.119pt}}
\multiput(444.00,121.17)(25.068,-13.000){2}{\rule{0.465pt}{0.400pt}}
\multiput(471.00,107.92)(1.052,-0.493){23}{\rule{0.931pt}{0.119pt}}
\multiput(471.00,108.17)(25.068,-13.000){2}{\rule{0.465pt}{0.400pt}}
\multiput(498.00,94.92)(0.974,-0.494){25}{\rule{0.871pt}{0.119pt}}
\multiput(498.00,95.17)(25.191,-14.000){2}{\rule{0.436pt}{0.400pt}}
\multiput(525.00,80.92)(1.052,-0.493){23}{\rule{0.931pt}{0.119pt}}
\multiput(525.00,81.17)(25.068,-13.000){2}{\rule{0.465pt}{0.400pt}}
\put(552,67.67){\rule{0.482pt}{0.400pt}}
\multiput(552.00,68.17)(1.000,-1.000){2}{\rule{0.241pt}{0.400pt}}
\end{picture}
\setlength{\unitlength}{0.240900pt}
\ifx\plotpoint\undefined\newsavebox{\plotpoint}\fi
\sbox{\plotpoint}{\rule[-0.200pt]{0.400pt}{0.400pt}}%
\begin{picture}(750,360)(0,0)
\font\gnuplot=cmr10 at 10pt
\gnuplot
\sbox{\plotpoint}{\rule[-0.200pt]{0.400pt}{0.400pt}}%
\put(176.0,135.0){\rule[-0.200pt]{122.859pt}{0.400pt}}
\put(176.0,68.0){\rule[-0.200pt]{4.818pt}{0.400pt}}
\put(154,68){\makebox(0,0)[r]{-20}}
\put(666.0,68.0){\rule[-0.200pt]{4.818pt}{0.400pt}}
\put(176.0,102.0){\rule[-0.200pt]{4.818pt}{0.400pt}}
\put(666.0,102.0){\rule[-0.200pt]{4.818pt}{0.400pt}}
\put(176.0,135.0){\rule[-0.200pt]{4.818pt}{0.400pt}}
\put(154,135){\makebox(0,0)[r]{0}}
\put(666.0,135.0){\rule[-0.200pt]{4.818pt}{0.400pt}}
\put(176.0,169.0){\rule[-0.200pt]{4.818pt}{0.400pt}}
\put(666.0,169.0){\rule[-0.200pt]{4.818pt}{0.400pt}}
\put(176.0,202.0){\rule[-0.200pt]{4.818pt}{0.400pt}}
\put(154,202){\makebox(0,0)[r]{20}}
\put(666.0,202.0){\rule[-0.200pt]{4.818pt}{0.400pt}}
\put(176.0,236.0){\rule[-0.200pt]{4.818pt}{0.400pt}}
\put(666.0,236.0){\rule[-0.200pt]{4.818pt}{0.400pt}}
\put(176.0,270.0){\rule[-0.200pt]{4.818pt}{0.400pt}}
\put(154,270){\makebox(0,0)[r]{40}}
\put(666.0,270.0){\rule[-0.200pt]{4.818pt}{0.400pt}}
\put(176.0,303.0){\rule[-0.200pt]{4.818pt}{0.400pt}}
\put(666.0,303.0){\rule[-0.200pt]{4.818pt}{0.400pt}}
\put(176.0,337.0){\rule[-0.200pt]{4.818pt}{0.400pt}}
\put(154,337){\makebox(0,0)[r]{60}}
\put(666.0,337.0){\rule[-0.200pt]{4.818pt}{0.400pt}}
\put(176.0,68.0){\rule[-0.200pt]{0.400pt}{4.818pt}}
\put(176,23){\makebox(0,0){9e-05}}
\put(176.0,317.0){\rule[-0.200pt]{0.400pt}{4.818pt}}
\put(337.0,68.0){\rule[-0.200pt]{0.400pt}{4.818pt}}
\put(337,23){\makebox(0,0){9.6e-05}}
\put(337.0,317.0){\rule[-0.200pt]{0.400pt}{4.818pt}}
\put(498.0,68.0){\rule[-0.200pt]{0.400pt}{4.818pt}}
\put(498.0,317.0){\rule[-0.200pt]{0.400pt}{4.818pt}}
\put(659.0,68.0){\rule[-0.200pt]{0.400pt}{4.818pt}}
\put(659,23){\makebox(0,0){0.000108}}
\put(659.0,317.0){\rule[-0.200pt]{0.400pt}{4.818pt}}
\put(176.0,68.0){\rule[-0.200pt]{122.859pt}{0.400pt}}
\put(686.0,68.0){\rule[-0.200pt]{0.400pt}{64.802pt}}
\put(176.0,337.0){\rule[-0.200pt]{122.859pt}{0.400pt}}
\put(256,159){\makebox(0,0)[l]{$\mu$}}
\put(271,228){\makebox(0,0)[r]{$\delta_{est}$}}
\put(256,313){\makebox(0,0)[l]{$\delta$}}
\put(605,297){\makebox(0,0)[l]{\it b}}
\put(176.0,68.0){\rule[-0.200pt]{0.400pt}{64.802pt}}
\put(176,135){\usebox{\plotpoint}}
\put(176.0,135.0){\rule[-0.200pt]{122.859pt}{0.400pt}}
\put(176,336){\usebox{\plotpoint}}
\multiput(176.00,334.92)(0.849,-0.494){29}{\rule{0.775pt}{0.119pt}}
\multiput(176.00,335.17)(25.391,-16.000){2}{\rule{0.388pt}{0.400pt}}
\multiput(203.00,318.92)(0.849,-0.494){29}{\rule{0.775pt}{0.119pt}}
\multiput(203.00,319.17)(25.391,-16.000){2}{\rule{0.388pt}{0.400pt}}
\multiput(230.00,302.92)(0.849,-0.494){29}{\rule{0.775pt}{0.119pt}}
\multiput(230.00,303.17)(25.391,-16.000){2}{\rule{0.388pt}{0.400pt}}
\multiput(257.00,286.92)(0.873,-0.494){27}{\rule{0.793pt}{0.119pt}}
\multiput(257.00,287.17)(24.353,-15.000){2}{\rule{0.397pt}{0.400pt}}
\multiput(283.00,271.92)(0.849,-0.494){29}{\rule{0.775pt}{0.119pt}}
\multiput(283.00,272.17)(25.391,-16.000){2}{\rule{0.388pt}{0.400pt}}
\multiput(310.00,255.92)(0.849,-0.494){29}{\rule{0.775pt}{0.119pt}}
\multiput(310.00,256.17)(25.391,-16.000){2}{\rule{0.388pt}{0.400pt}}
\multiput(337.00,239.92)(0.908,-0.494){27}{\rule{0.820pt}{0.119pt}}
\multiput(337.00,240.17)(25.298,-15.000){2}{\rule{0.410pt}{0.400pt}}
\multiput(364.00,224.92)(0.849,-0.494){29}{\rule{0.775pt}{0.119pt}}
\multiput(364.00,225.17)(25.391,-16.000){2}{\rule{0.388pt}{0.400pt}}
\multiput(391.00,208.92)(0.908,-0.494){27}{\rule{0.820pt}{0.119pt}}
\multiput(391.00,209.17)(25.298,-15.000){2}{\rule{0.410pt}{0.400pt}}
\multiput(418.00,193.92)(0.938,-0.494){25}{\rule{0.843pt}{0.119pt}}
\multiput(418.00,194.17)(24.251,-14.000){2}{\rule{0.421pt}{0.400pt}}
\multiput(444.00,179.92)(0.974,-0.494){25}{\rule{0.871pt}{0.119pt}}
\multiput(444.00,180.17)(25.191,-14.000){2}{\rule{0.436pt}{0.400pt}}
\multiput(471.00,165.92)(1.251,-0.492){19}{\rule{1.082pt}{0.118pt}}
\multiput(471.00,166.17)(24.755,-11.000){2}{\rule{0.541pt}{0.400pt}}
\multiput(498.00,154.93)(1.543,-0.489){15}{\rule{1.300pt}{0.118pt}}
\multiput(498.00,155.17)(24.302,-9.000){2}{\rule{0.650pt}{0.400pt}}
\multiput(525.00,145.93)(2.389,-0.482){9}{\rule{1.900pt}{0.116pt}}
\multiput(525.00,146.17)(23.056,-6.000){2}{\rule{0.950pt}{0.400pt}}
\multiput(552.00,139.95)(5.820,-0.447){3}{\rule{3.700pt}{0.108pt}}
\multiput(552.00,140.17)(19.320,-3.000){2}{\rule{1.850pt}{0.400pt}}
\put(579,136.17){\rule{5.300pt}{0.400pt}}
\multiput(579.00,137.17)(15.000,-2.000){2}{\rule{2.650pt}{0.400pt}}
\put(605,134.67){\rule{6.504pt}{0.400pt}}
\multiput(605.00,135.17)(13.500,-1.000){2}{\rule{3.252pt}{0.400pt}}
\put(632.0,135.0){\rule[-0.200pt]{13.009pt}{0.400pt}}
\put(176,299){\usebox{\plotpoint}}
\multiput(176.00,297.92)(0.849,-0.494){29}{\rule{0.775pt}{0.119pt}}
\multiput(176.00,298.17)(25.391,-16.000){2}{\rule{0.388pt}{0.400pt}}
\multiput(203.00,281.92)(0.798,-0.495){31}{\rule{0.735pt}{0.119pt}}
\multiput(203.00,282.17)(25.474,-17.000){2}{\rule{0.368pt}{0.400pt}}
\multiput(230.00,264.92)(0.849,-0.494){29}{\rule{0.775pt}{0.119pt}}
\multiput(230.00,265.17)(25.391,-16.000){2}{\rule{0.388pt}{0.400pt}}
\multiput(257.00,248.92)(0.817,-0.494){29}{\rule{0.750pt}{0.119pt}}
\multiput(257.00,249.17)(24.443,-16.000){2}{\rule{0.375pt}{0.400pt}}
\multiput(283.00,232.92)(0.798,-0.495){31}{\rule{0.735pt}{0.119pt}}
\multiput(283.00,233.17)(25.474,-17.000){2}{\rule{0.368pt}{0.400pt}}
\multiput(310.00,215.92)(0.849,-0.494){29}{\rule{0.775pt}{0.119pt}}
\multiput(310.00,216.17)(25.391,-16.000){2}{\rule{0.388pt}{0.400pt}}
\multiput(337.00,199.92)(0.798,-0.495){31}{\rule{0.735pt}{0.119pt}}
\multiput(337.00,200.17)(25.474,-17.000){2}{\rule{0.368pt}{0.400pt}}
\multiput(364.00,182.92)(0.849,-0.494){29}{\rule{0.775pt}{0.119pt}}
\multiput(364.00,183.17)(25.391,-16.000){2}{\rule{0.388pt}{0.400pt}}
\multiput(391.00,166.92)(0.849,-0.494){29}{\rule{0.775pt}{0.119pt}}
\multiput(391.00,167.17)(25.391,-16.000){2}{\rule{0.388pt}{0.400pt}}
\multiput(418.00,150.92)(0.768,-0.495){31}{\rule{0.712pt}{0.119pt}}
\multiput(418.00,151.17)(24.523,-17.000){2}{\rule{0.356pt}{0.400pt}}
\multiput(444.00,133.92)(0.849,-0.494){29}{\rule{0.775pt}{0.119pt}}
\multiput(444.00,134.17)(25.391,-16.000){2}{\rule{0.388pt}{0.400pt}}
\multiput(471.00,117.92)(0.798,-0.495){31}{\rule{0.735pt}{0.119pt}}
\multiput(471.00,118.17)(25.474,-17.000){2}{\rule{0.368pt}{0.400pt}}
\multiput(498.00,100.92)(0.849,-0.494){29}{\rule{0.775pt}{0.119pt}}
\multiput(498.00,101.17)(25.391,-16.000){2}{\rule{0.388pt}{0.400pt}}
\multiput(525.00,84.92)(0.849,-0.494){29}{\rule{0.775pt}{0.119pt}}
\multiput(525.00,85.17)(25.391,-16.000){2}{\rule{0.388pt}{0.400pt}}
\put(552,68.17){\rule{0.482pt}{0.400pt}}
\multiput(552.00,69.17)(1.000,-2.000){2}{\rule{0.241pt}{0.400pt}}
\end{picture}
\caption{\label{Graphs}Concentration dependences of selectivity for the examples
of Table \ref{Table}. Concentration (x-axis) is given in M.
The $N$, $N_0$ values in {\it a} and {\it b} correspond
to the
first and second rows of Table \ref{Table}, respectively. The $\delta_{est}$
corresponds to the right hand side of the inequality (\ref{ineq1}).}
\end{figure}
Two examples satisfying this constraints are shown in the Table \ref{Table}.
Concentration dependencies of $\mu$, $\delta$, and the estimate (\ref{ineq1})
are shown in Fig.\ref{Graphs}. A short segment of the trajectory $n(t)$
modelled on PC is shown in Fig.\ref{Trajec}.

 \section{Conclusions and discussion}

In this paper, selectivity of chemical sensor is compared with that of
its primary receptors (adsorbing sites).
The sensor is expected to be a small one, in which the main source of noise is
due to the adsorption-desorption fluctuations. In the sensor considered, the
signal from the primary sensing unit is immediately subjected to the amplitude
discrimination defined in the Introduction,
and obtained piecewise-constant signal ($L(t)$ in
Fig.\ref{ThD}) is averaged over a time window. The averaged signal ($S$ in
Fig.\ref{ThD}) is taken as the output of whole sensor.

It is concluded that
selectivity of this sensor can be much better than that of its primary
receptors. The effect may be expected in a limited range of concentrations of
analytes, which depends on the threshold level. For high concentrations the
selectivity falls to that of the primary receptors (Fig.\ref{Graphs}), and
for low ones the output signal will be too small even for more affine
analyte. The best situation is expected
when the mean number of bound receptors is just
below the threshold one, and the threshold is
frequently crossed due to the presence
of fluctuations. Thus, in practical realization a possibility of tuneable
threshold should be considered.

Usually, noise in sensory devices is taken as unfavorable factor\footnote{but
see \cite{Smulko}, where some characteristics of noise are employed for
discriminating purposes.}. In this consideration, the presence of noise looks
like factor improving the sensor performance. But with
the ideal threshold unit
in hands much can be done even without noise.
Expect that the noise is
initially averaged out either by spatial averaging (choosing big primary unit
with large $N$), or by temporal averaging (interchanging TAU with ThU in
Fig.\ref{ThD}). The averaged signals for the $A_1$, $A_2$ can be very close
(see Eq. \ref {linprop}), but the ideal ThU with tunable threshold will be able
to discriminate perfectly between them. Thus, even if the fluctuations in
this sensor are made working, the answer what is better to do first
for the practical purposes: the amplitude discrimination, or
temporal averaging, depends on physical parameters of the
environment in which the sensor operates, and on physical characteristics of
the sensor itself, including intensity of noises other than the
adsorption-desorption one.  Interesting, in natural olfactory systems, a kind
of amplitude discrimination is made immediately after the primary reception
\cite{VidBS,Rospars}. Also in those systems the threshold is tunable due to
adaptation of individual neurons.\bigskip

{\small
\textit{Acknowledgments}. This work was supported by the Programs of Basic Research of the Department of Physics and Astronomy of the National Academy of Sciences of Ukraine ``Mathematical models of nonequilibrium processes in open systems'', № 0120U100857, and ``Noise-induced dynamics and correlations in nonequilibrium systems'',
№ 0120U101347.

\end{document}